\documentclass[aps, prd, preprintnumbers, floatfix, showpacs, showkeys, nofootinbib, 10pt]{revtex4-1}

\usepackage[dvips]{graphics}
\usepackage{amsmath,amsfonts,bm}
\usepackage{amssymb}
\usepackage{amsthm}
\usepackage{epsfig,bm}
\usepackage{feynmp}
\usepackage{graphicx,comment}
\usepackage{dcolumn,color}
\usepackage{mathtools}
\usepackage{fancyhdr}
\usepackage{bm}
\usepackage{cancel}
\usepackage{slashed}
\usepackage{verbatim}
\usepackage{listings}
\usepackage{graphicx,latexsym}
\usepackage{rotating}
\usepackage{color}
\usepackage{float}
\usepackage{enumerate}
\usepackage{array}
\usepackage{tabularx}
\usepackage{longtable}
\usepackage{booktabs}

    \def\Im{\mathop{\rm Im}}

\def\beq{\begin{equation}}
\def\eeq{\end{equation}}
\def\bea{\begin{eqnarray}}
\def\eea{\end{eqnarray}}

\newcommand{\be}{\begin{equation}}
\newcommand{\ee}{\end{equation}}
\newcommand{\bear}{\begin{eqnarray}}
\newcommand{\eear}{\end{eqnarray}}
\newcommand{\ba}{\begin{array}}
\newcommand{\ea}{\end{array}}

\begin{document}
\preprint{\today}


\title{A Second Look at String-Inspired Models for Proton-Proton Scattering via Pomeron Exchange}

\author{Ziyi~Hu$^1$, Brian~Maddock$^2$, Nelia~Mann$^1$}
\affiliation{$^1$ Union College, Schenectady NY, 12308, USA}
\affiliation{$^2$ University of Cincinnati, Cincinnati OH, 45220, USA}


\begin{abstract}
We re-examine a string dual model for elastic proton-proton scattering via Pomeron exchange.  We argue that the method of ``Reggeizing'' a propagator to take into account an entire trajectory of exchanged particles can be generalized, in particular by modifying the value of the mass-shell parameter in the model.  We then fit the generalized model to scattering data at large $s$ and small $t$.  The fitting results are inconclusive, but suggest that a better fit might be obtained by allowing the mass-shell to vary.  The model fits the data equally well (roughly) for a wide range of values of the mass-shell parameter, but the other fitting parameters (the slope and intercept of the Regge trajectory, and the coupling constant and dipole mass from the proton-proton-glueball coupling) are then inconsistent with what we expect.  On the other hand, using the traditional method of Reggeization generates a weaker fit, but the other parameters obtain more physically reasonable values.  In analyzing the fitting results, we also found that our model is more consistent with the $\sqrt{s} = 1800$ GeV data coming from the E710 experiment than that coming from the CDF experiment, and that our model has the greatest discrepancy with the data in the range $0.5 \ \mathrm{GeV}^2 < |t| < 0.6 \ \mathrm{GeV}^2$, suggesting that the transition from soft Pomeron to hard Pomeron may occur closer to $t = -0.5 \ \mathrm{GeV}^2$ rather than $t = -0.6 \ \mathrm{GeV}^2$ as previously thought.
\end{abstract}



\maketitle


\section{Introduction}

Elastic hadron scattering in the Regge regime (high center-of-mass energy and small scattering angle) has long been understood to have particularly interesting features.  Written in terms of the Mandelstam variables $s$ and $t$, the scattering amplitude scales as $s^{\alpha(t)}$, where $\alpha(t)$ is a linear function known as a Regge trajectory.  For positive values of $t$, the amplitudes have regularly spaced poles associated to mesons of masses $m^2 = t$, and spins $J =   \alpha(m^2)$.  The interpretation of this is that the full scattering amplitude can be thought of as an infinite sum of amplitudes associated with the exchanges of the mesons lying along the trajectory \cite{earlyRegge}.

Analysis of this type of behavior in baryon and meson scattering is tied to the earliest work in string theory: the Veneziano amplitude originally written down to model pion scattering was later shown to arise naturally from the scattering of open strings.  In general, string amplitudes are known to have the same scaling behaviors in the Regge regime, and the same dependence on a linear Regge trajectory.  This Regge trajectory also represents the linear relationship between mass squared and spin for spin states.  However, string amplitudes can only (thus far) be calculated for simple cases such as 26-d flat space Bosonic string theory.  For these cases, the string states and Regge trajectory parameters do not correspond well to physical mesons.

More than a decade ago now, the idea of a relationship between QCD and string theory gained new traction with the proposal that QCD might be dual to string theory living in a curved, 5-dimensional space.  In this scenario we have mesons mapping onto open strings, glueballs mapping onto closed strings, and baryons mapping onto D-branes.  Various toy models for such a duality have been proposed, including the Sakai-Sugimoto model in which stacks of D-branes are placed in a background space to generate curvature, as well as soft-wall and hard-wall models \cite{Sakai:2004cn}.  Although such models have limitations, they typically agree reasonably well with masses and coupling constants coming from both experimental results and lattice QCD calculations.  

At very high center-of-mass energies proton-proton scattering and proton-antiproton scattering data suggest a trajectory not consistent with that of any known mesons: possessing a higher intercept and consisting of particles of even spin and vacuum quantum numbers (so that proton-proton data is identical to proton-antiproton data).  We will adapt the common interpretation of this, that these processes are mediated by a single trajectory of glueballs known as the Pomeron \cite{CGM}.  However, direct experimental confirmation of the glueballs involved does not exist, though they are observed in lattice QCD calculations \cite{lattice}.  In addition, it is known that at the highest energies, total cross sections must obey the Froissart-Martin bound and grow no faster than $(\ln s)^2$, implying something more complicated than single Pomeron exchange.  Whether scattering data existing today (up to energies reached at the LHC) is at high enough energies to be affected by this is still an open question, though we will assume that it is not.

Interpretation of the Pomeron trajectory is complicated by the fact that it has a significantly different behavior in the hard scattering regime $|t| > \Lambda^2_{\mathrm{QCD}}$, where it ought to be associated to a sum over perturbative QCD processes involving gluon exchange, than it does in the soft scattering regime $0 < |t| < \Lambda^2_{\mathrm{QCD}}$, where the exchange of bound glueball states makes more sense.  How a single trajectory could have these different behaviors can be understood within the string dual picture: in \cite{bpstetc}, it was shown that the radius of curvature of the 5th dimension generates an energy scale that can be mapped onto $\Lambda_{\mathrm{QCD}}$, such that the closed string trajectory would have different behaviors for low and high energies as compared to this scale.  Work has also been done analyzing the structure of the Pomeron trajectory in soft- or hard-wall models for holographic QCD, and in analyzing the Pomeron trajectory in backgrounds with an arbitrary number of dimensions \cite{topdown}.  In this paper we will restrict our attention to the soft Pomeron regime, and assume the trajectory is linear.

Building a string-dual model to explain proton-proton scattering via Pomeron exchange in the Regge regime is a project of real interest in this conversation.  However, calculations within toy dual models are generally restricted to the supergravity limit, corresponding to low energy QCD processes, so additional tools are necessary to extend the usefulness of AdS/QCD outside this regime.  In \cite{DHM}, an assumption was suggested that the main structures of string scattering amplitudes in flat space (in this case the Virasoro-Shapiro amplitude) would also apply to the scattering amplitudes in weakly curved spacetimes, but with the defining parameters of the Regge trajectory modified by the curvature.  Furthermore, it was proposed that the coupling constants appearing in these amplitudes would be the same as those calculable in the low-energy limit.  This leads to a hybrid approach for modeling scattering processes in the Regge regime: coupling constants are determined in the supergravity limit using a toy model, and low energy scattering cross sections are determined using them.  Then, the propagators in these cross sections are ``Reggeized'' using a modified version of a string scattering amplitude, with the Regge trajectory parameters chosen to agree with the physical trajectory.  This basic procedure was extended in \cite{ADHM, ADM, IRS} to apply to central-production processes, with the Reggeization based on 5-string amplitudes.  

In general, this method agrees reasonably well with experimental results in some respects but has significant discrepancies in others \cite{ADM}.  It is possible these discrepancies arise due to the limitations of using a toy dual model to generate the low energy coupling constants; recent work has suggested for example that the Sakai-Sugimoto model systematically underestimates coupling constant values \cite{low}.  However, it is also possible that the Reggeization procedure used is not ideal: perhaps the assumptions made about string amplitudes in a weakly curved background are not correct.  In this paper we seek to examine this latter concern by revisiting the Reggeization procedure for elastic proton-proton scattering via Pomeron exchange and attempting to introduce generalizations where possible, while still maintaining the phenomenologically desirable features of the amplitude's behavior.  In particular, the Reggeization procedure of \cite{DHM} introduces a parameter $\chi$ that arises from the mass-shell condition for the external particles and that depends on the Pomeron trajectory.  However, we will show that our generalizations allow for other values of $\chi$.  We will also examine the issue by comparing the generalized model with real data, and allowing $\chi$ to be a fitting parameter.  The results of the fitting procedure are ambiguous, but they suggest that the value of $\chi$ used previously may not be the best choice for agreement with the data.

In section \ref{PomeronReview}, we will review what is known about proton-proton scattering via the exchange of a single Pomeron trajectory, and we will identify the key desirable features to look for in a string amplitude designed to model this process.  In section \ref{mod1}, we will review the Reggeization procedure of \cite{DHM}, and introduce generalizations consistent with these features.  We will show that these generalizations amount to allowing the value of the mass-shell parameter $\chi$ to change.  In section \ref{mod2}, we will show a second, related Reggeization procedure, leading again to a different choice of $\chi$.  In section \ref{fitting}, we compare the model to scattering data, allowing the value of $\chi$ to be a fitting parameter. In section \ref{Conclusion}, we offer discussion and conclusions.

\section{\label{PomeronReview} Reviewing Pomeron Exchange in Proton-Proton Scattering}

In this section we will (briefly) review some of the essentials of Regge theory.  There are no new results presented here; our goal is to establish what the phenomenological requirements should be for an amplitude designed to model pomeron exchange in proton-proton scattering.  

\begin{figure}
\begin{center}
\resizebox{4in}{!}{\includegraphics{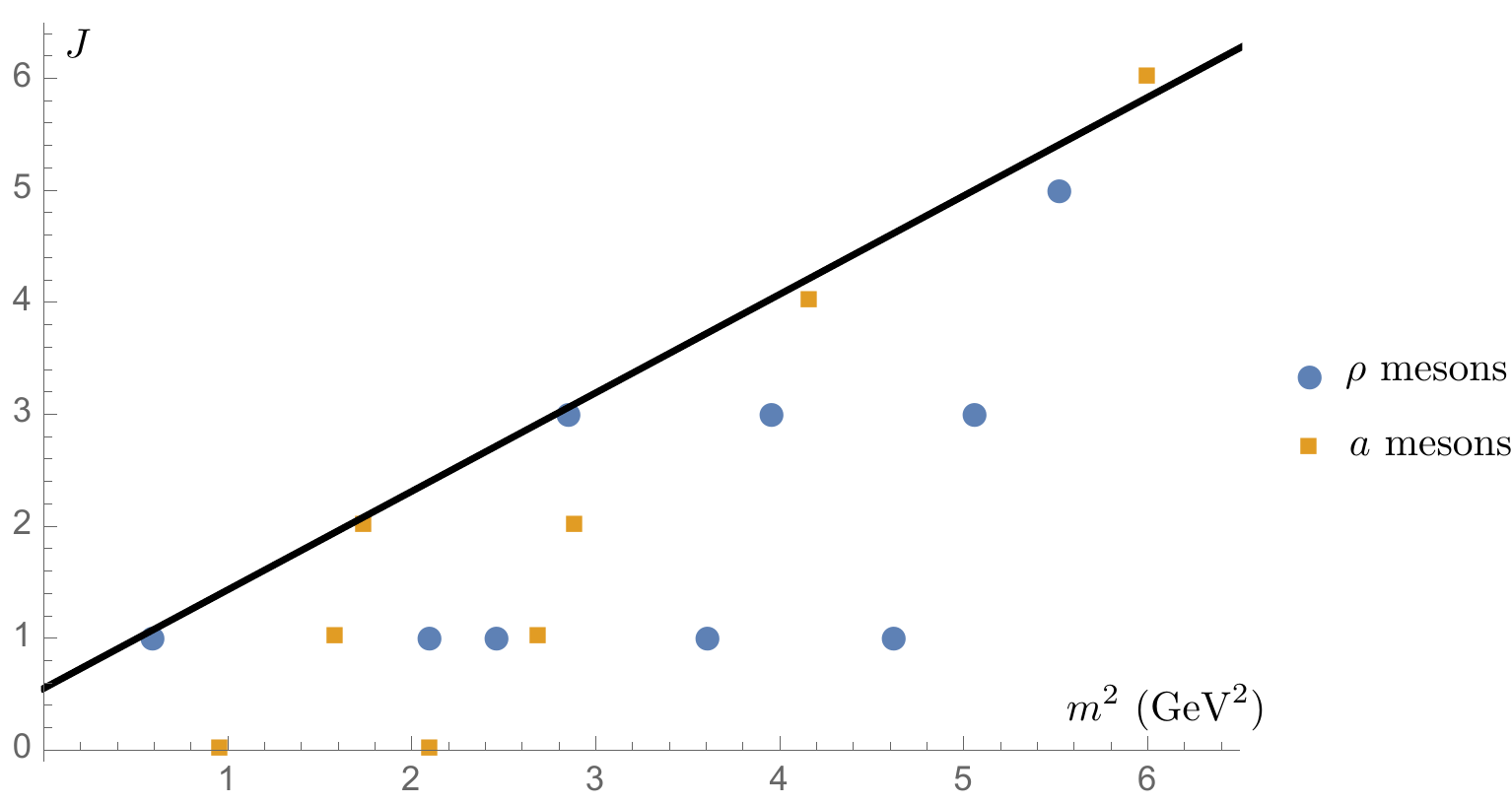}}
\caption{\label{mesontraj} A Regge plot of $\rho$ and $a$ mesons, showing the leading trajectory \cite{PDG}.}
\end{center}
\end{figure}

Consider elastic proton-proton scattering or proton-antiproton scattering expressed in terms of the standard Mandelstam variables $s$ and $t$, in the Regge limit where $s \gg t$.  We can describe both the differential and total cross sections in terms of an amplitude $\mathcal{A}(s, t)$ as
\beq
\sigma_{\mathrm{tot}} = \frac{1}{s} \, \Im\mathcal{A}(s, 0), \hspace{1in} \frac{d\sigma}{dt} = \frac{1}{16\pi s^2}\left|\mathcal{A}(s, t)\right|^2 \, .
\eeq
This scattering process occurs via the exchanges of families of particles that lie along Regge trajectories \cite{earlyRegge}.  Mesons, baryons, and glueballs form patterns where there is a linear relationship between the spin of a ``family member'' and its mass squared:
\beq
J = \alpha_0 + \alpha' m_J^2 \, ,
\eeq
so that we can define the linear Regge trajectory function as
\beq
\alpha(x) = \alpha_0 + \alpha' x, \hspace{1in} J = \alpha(m_J^2) \, .
\eeq
For example, consider the $\rho$ and $a$ mesons, shown in FIG. \ref{mesontraj}.  There is a leading linear trajectory of particles with the smallest mass for a given spin, with $\alpha_0 \approx 0.53$ and $\alpha' \approx 0.88 \ \mathrm{GeV}^{-2}$.  

The scattering amplitude, differential cross section, and total cross section for the exchange of this leading trajectory (which will dominate over the ``daughter trajectories'' in the Regge limit) are known to take the generic form
\beq
\mathcal{A}(s, t) = \beta(t) \left(\alpha' s\right)^{\alpha(t)}, \hspace{.5in} \frac{d\sigma}{dt} =  \frac{\alpha'^2 |\beta(t)|^2}{16\pi}\left(\alpha' s\right)^{2\alpha(t) - 2}, \hspace{.5in} \sigma_{\mathrm{tot}} = \frac{\Im \beta(0)}{s} \, \left(\alpha' s\right)^{\alpha_0 - 1} \, .
\eeq
Both the pole structure associated with the exchange of a Regge trajectory of particles and the characteristic Regge limit scaling behavior are consistent with the Veneziano amplitude, which can be written in its original form as
\beq
\mathcal{A}^{\mathrm{Ven}}_{\{n, m, p\}}(s, t) = \frac{\Gamma[n - \alpha(s)]\Gamma[m - \alpha(t)]}{\Gamma[p - \alpha(s) - \alpha(t)]} \, .
\eeq
This (famously) is also a structure that arises in the scattering of open strings in 26-D flat spacetime, where the amplitude for the scattering of four tachyonic scalar string states would take a crossing-symmetric form generated by a sum of three such amplitudes, with $n = m = p = 0$.  However, the actual Regge trajectories arising do not compare well with the known masses and spins of vector mesons.  (For example, there is no tachyonic vector meson.)

\begin{figure}
\begin{center}
\resizebox{4in}{!}{\includegraphics{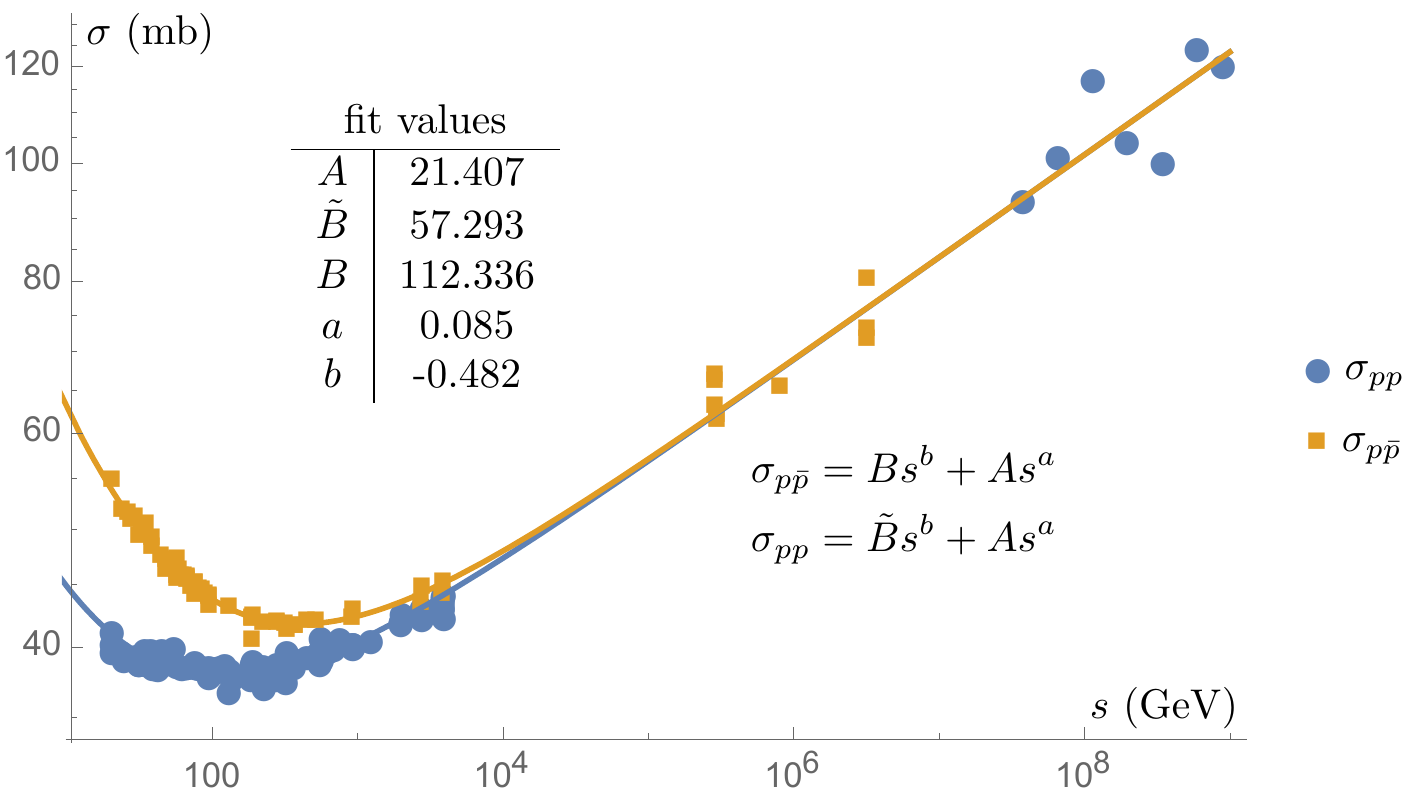}}
\caption{\label{totalcross} The total cross sections for proton-proton scattering and proton-antiproton scattering \cite{PDG}.}
\end{center}
\end{figure}

At very high center-of-mass energies, there is significant evidence that both proton-proton and proton-antiproton scattering are dominated by what is known as the Pomeron trajectory.  Consider for example the total cross sections of proton-proton scattering and proton-antiproton scattering, which are plotted as a function of $s$ in FIG. \ref{totalcross}.  We can see that this data is well fit by assuming that two separate Regge trajectories with two separate intercepts contribute to the scattering process.  At lower energies the process is controlled by a term that corresponds well with the known parameters of the $\rho-a$ trajectory.  Note that both proton-proton and proton-antiproton scattering have such a contribution, but that the scale of this trajectory's contribution to proton-proton scattering is somewhat smaller, since the alternating even and odd spins have partially canceling effects.  

For very high values of $s$, the leading term suggests a Regge trajectory with intercept around $1.08$, which is larger than that of any known meson trajectory.  Furthermore, the contribution from this term is equal for both proton-proton scattering and proton-antiproton scattering.  This suggests the trajectory is made up of particles with vacuum quantum numbers, and in particular that only even-spin particles appear on it.  We have taken the common point of view that this is associated with a single glueball trajectory: the Pomeron \cite{CGM}.

Based on this evidence, we can say that a phenomenologically valid Reggeization procedure for Pomeron exchange should begin with an amplitude that is fully crossing symmetric, since we know Pomeron exchange treats particles and anti-particles identically.  Also, the pole structure should correspond to the exchanges of (even spin) glueballs on the Pomeron trajectory: if we assume
\beq
J = \alpha_{g0} + \alpha_{g}'m_{g, J}^2 = \alpha_g(m_{g, J}^2) \, ,
\eeq
is the trajectory of glueballs, with lowest lying state having spin $2$, we should have a pole at every $\alpha_g(t) = 2, 4, 6, 8, \dots$, with the structure
\beq
\mathcal{A}_{t \approx m_{g, J}^2} \approx \frac{P_{J}(s)}{t - m_{g, J}^2} \, ,
\eeq
where $P_J(s)$ is a polynomial of degree $J$ in $s$.  Finally, it should have the correct Regge behavior:
\beq
\mathcal{A}_{\mathrm{Regge}} \approx \beta(t) \left(\alpha_g's\right)^{\alpha_g(t)} \, .
\eeq

Glueballs ought to be dual to closed strings in some hyperbolically curved background.  The Reggeization method of \cite{DHM}, further elaborated on in \cite{ADHM} and \cite{ADM}, takes the approach of staying as close to the flat-space string theory amplitude (the Virasoro-Shapiro amplitude) as possible, making only the minimal modifications necessary to meet all of the requirements above.  However, as we will see in the next two sections, other modifications are possible.  It is then possible to use fitting to data to determine which modification scheme is actually in the best agreement with reality, and in particular how the original ``minimal modification'' scheme fares in comparison to others.  

\section{\label{mod1} A Variation on the Known Method of Modifying the Virasoro-Shapiro Amplitude}

The standard approach to determining the Reggeization procedure for a propagator, first presented in \cite{DHM}, is to begin with the Virasoro-Shapiro amplitude written in a form with manifest crossing symmetry, and then modify both the Regge trajectory parameters and the mass shell condition such that the amplitude has the correct pole structure, before taking the Regge limit.  The Reggeization procedure can then be read off from a comparison between the pole expansion and the Regge limit.  Here we will review this procedure while allowing for some generalization, in order to examine how unique the result is.  

To begin, consider closed Bosonic strings in 26-D flat space.  Here, the spectrum of string states lying on the leading Regge trajectory have even spins $J$ and masses $m_J$ satisfying
\beq
J = 2 + \frac{\alpha'}{2}m_J^2 = a_c(m_J^2) \, .
\eeq
where $a_c(x)$ is the Regge trajectory.  The lowest lying states are scalar tachyons with masses satisfying $m_T^2 = -\frac{4}{\alpha'}$.  The tree-level scattering amplitude for four external tachyonic states gives us the well-known Virasoro-Shapiro amplitude, which can be written as
\beq
\label{eqn:VSoriginal}
\mathcal{A}_{c}(s, t, u) = \frac{C \, \pi \, \Gamma\left(-\frac{a_c(s)}{2}\right) \, \Gamma\left(-\frac{a_c(t)}{2}\right) \, \Gamma\left(-\frac{a_c(u)}{2}\right)}{\Gamma\left(-\frac{a_c(s)}{2} - \frac{a_c(t)}{2}\right) \, \Gamma\left(-\frac{a_c(t)}{2} - \frac{a_c(u)}{2}\right) \, \Gamma\left(-\frac{a_c(s)}{2} - \frac{a_c(u)}{2}\right)} \, ,
\eeq
where the mass-shell condition on Mandelstam variables is
\beq
\label{eqn:BSMS}
s + t + u = -\frac{16}{\alpha'} \, , \hspace{1in} a_c(s) + a_c(t) + a_c(u) = -2 \, .
\eeq
Note that in this form the crossing symmetry of the amplitude is manifest.  If we rewrite this in terms of just the independent variables $s$ and $t$ we obtain
\beq
\label{eqn:BSVSts}
\mathcal{A}_{c}(s, t) = \frac{C \, \pi \, \Gamma\left(-\frac{a_c(s)}{2}\right) \, \Gamma\left(-\frac{a_c(t)}{2}\right) \, \Gamma\left(1 + \frac{a_c(s)}{2} + \frac{a_c(t)}{2}\right)}{\Gamma\left(-\frac{a_c(t)}{2} - \frac{a_c(s)}{2}\right) \, \Gamma\left(1 +  \frac{a_c(s)}{2}\right) \, \Gamma\left(1 + \frac{a_c(t)}{2}\right)} \, .
\eeq
This amplitude has a pole for each $t = m_J^2$.  If we expand around a pole corresponding to the exchange of a particle of spin $J$, we obtain
\beq
\mathcal{A}_{c, \, t \approx m_J^2}(s, t) \approx \frac{4C\pi \, e^{-i\pi\left(\frac{J}{2} + 1\right)} \, P_J\left(\frac{\alpha' s}{4}\right)}{\alpha'\left[\Gamma\left(\frac{J}{2} + 1\right)\right]^2 \, (t - m_J^2) } \, ,
\eeq
where $P_J\left(\frac{\alpha' s}{4}\right)$ is a polynomial in $s$ of degree $J$ such that $P_J\left(\frac{\alpha' s}{4}\right) = \left(\frac{\alpha' s}{4}\right)^J + \cdots$.  On the other hand, if we examine the Regge limit of this amplitude, we obtain
\beq
\mathcal{A}_{c, \, \mathrm{Regge}}(s, t) \approx \frac{C\pi \, \Gamma\left(-\frac{a_c(t)}{2}\right)}{\Gamma\left(1 + \frac{a_c(t)}{2}\right)} \, e^{-\frac{i\pi a_c(t)}{2}}\left(\frac{\alpha' s}{4}\right)^{a_c(t)} \, .
\eeq
All of these behaviors are just as expected:  the poles are in the correct locations and have the right residues to correspond to exchanges of particles on the trajectory of open string states, and in the Regge limit we have correct scaling behavior with $s$, also associated with the trajectory.  The Reggeization procedure assumes we are working in a low energy limit with a scattering process involving the exchange of the lowest lying particle on a trajectory (in this case, a tachyon), and that we can use that amplitude or cross section for the entire trajectory of exchanged particles in the Regge limit, by simply replacing the propagator with a Reggeized version.  In this case, that replacement rule would be
\beq
\frac{1}{t - m_T^2} \hspace{.25in} \rightarrow \hspace{.25in} -\frac{\alpha' \Gamma\left(-\frac{a_c(t)}{2}\right)}{4\Gamma\left(1 + \frac{a_c(t)}{2}\right)} \, e^{-\frac{i\pi a_c(t)}{2}} \, \left(\frac{\alpha' s}{4}\right)^{a_c(t)} \, .
\eeq

However, as we know, this rule lacks some essential features: the particles being scattered are not protons, and those being exchanged are not glueballs.  We therefore want to make alterations to this rule so that it corresponds to the scattering of physical particles, while retaining the many desirable features it already has.  The procedure of \cite{DHM} involves replacing the dependence on the closed string Regge trajectory with an unknown linear function $A(x)$ (such that the crossing symmetry is maintained), and then relating this back to the physical glueball trajectory by requiring that the new amplitude has the correct pole structure.  However, from a phenomenological point of view, we can actually introduce two separate unknown linear functions, which we will call $A(x)$ and $\tilde{A}(x)$, while still maintaining the desired crossing symmetry.  We also multiply by an unknown kinematic factor $F(s, t, u)$.  In the method of \cite{DHM}, this is eventually chosen to be $F(s, t, u) = s^2 + t^2 + u^2$, which allows for the residues of the poles to have the correct scaling with $s$.  Again, we will leave this arbitrary for now.  Thus our starting place is
\beq
\label{eqn:VSgxs}
\mathcal{A}_g(s, t, u) = \frac{C \, \pi \, \Gamma\left(-A(s)\right)\Gamma\left(-A(t)\right)\Gamma\left(-A(u)\right)}{\Gamma\left(-\tilde{A}(s) - \tilde{A}(t)\right)\Gamma\left(-\tilde{A}(t) - \tilde{A}(u)\right)\Gamma\left(-\tilde{A}(s) - \tilde{A}(u)\right)} \, F(s, t, u) \, .
\eeq

Next we want to examine pole structure and Regge limit behavior, but for both of these we need to rewrite our amplitude in term of just the independent Mandelstam variables $t$ and $s$.  Doing so introduces new parameters, 
\beq
A(s) + A(t) + A(u) = \chi, \hspace{1in} \tilde{A}(s) + \tilde{A}(t) + \tilde{A}(u) = \tilde{\chi} \, .
\eeq
Note that these must be constants, assuming both $A(x)$ and $\tilde{A}(x)$ are linear, and are determined by the mass shell condition we impose in assuming the external particles are protons, which is
\beq
\label{eqn:msprotons}
s + t + u = 4m_p^2 \, .
\eeq
This gives
\beq
\label{eqn:VSgts}
\mathcal{A}_g(s, t) = \frac{C \, \pi \, \Gamma\left(-A(s)\right)\Gamma\left(-A(t)\right)\Gamma\left(A(s) + A(t) - \chi\right)}{\Gamma\left(-\tilde{A}(s) - \tilde{A}(t)\right)\Gamma\left(\tilde{A}(s) - \tilde{\chi}\right)\Gamma\left(\tilde{A}(t) - \tilde{\chi}\right)} \, F(s, t, 4m_p^2 - s - t) \, .
\eeq
This amplitude has a pole at every value $A(t) = n$, where $n$ is an integer, which we want to correspond to the masses of the physical glueballs.  (At this stage we need to require that the function $F$ neither cancels any of these poles nor introduces new ones into the amplitude.)  Suppose we define $\alpha_g(x)$ as the Pomeron trajectory, with
\beq
\alpha_g(x) = \alpha_{g0} + \alpha_g'x \, .
\eeq
In order to agree with what we believe about the physical Pomeron, we want the lowest lying particle on this trajectory to have spin $2$, the next spin $4$, and so on.  This implies
\beq
A(x) = \frac{\alpha_g(x)}{2} - 1 \, .
\eeq
Note that if we initially allowed for further generalization by making $A(x)$ an arbitrary function instead of requiring that it be linear, we would be led to the same place at this stage.  Furthermore, our definition of $\chi$ now agrees with that in \cite{DHM}\footnote{We could also convert this to the notation of \cite{ADM}, which uses $\alpha_g(s) + \alpha_g(t) + \alpha_g(u) = \chi_g$, so that $\chi = \frac{\chi_g}{2} - 3$.}.  Next we examine the expansion near one of these poles, to ensure that we obtain the correct scaling of the residue with $s$.  This will give

\beq
\mathcal{A}_{g, t \approx m_{g, J}^2} \approx
\eeq
\nopagebreak
$$
 \frac{2C\pi e^{-\frac{i\pi J}{2}}}{\alpha_g'\Gamma\left(\frac{J}{2}\right)\Gamma\left(\tilde{A}(m_{g, J}^2) - \tilde{\chi}\right)(t - m_{g, J}^2)}\left[\frac{\Gamma\left(1 - \frac{\alpha_g(s)}{2}\right)}{\Gamma\left(-\tilde{A}(s) - \tilde{A}(m_{g, J}^2)\right)}\right]\left[\frac{\Gamma\left(\frac{\alpha_g(s)}{2} + \frac{J}{2} - 2 - \chi\right)}{\Gamma\left(\tilde{A}(s) - \tilde{\chi}\right)}\right]F(s, m_{g, J}^2, 4m_p^2 - s - m_{g, J}^2) \, .
$$
In the usual scheme, $A(x) = \tilde{A}(x)$, and each of the ratios of Gamma functions above yields a polynomial in $s$.  Since this is an essential part of the residue structure, we would like to maintain this behavior.  However, we see that we can do so provided $A(x)$ and $\tilde{A}(x)$ differ by at most a half-integer.  We will therefore be content with the looser requirement that
\beq
\tilde{A}(x) = A(x) + \frac{k}{2}, 
\hspace{.75in} \tilde{\chi} = \chi + \frac{3k}{2}, \hspace{.75in} k \in \mathbb{Z} \, .
\eeq
This then gives
\beq
\mathcal{A}_{g, t \approx m_{g, J}^2} \approx
 \frac{2C\pi e^{-\frac{i\pi J}{2}}P_{J + 2k - 2}\left(\frac{\alpha_g' s}{2}\right)}{\alpha_g'\Gamma\left(\frac{J}{2}\right)\Gamma\left(\frac{J}{2} - 1 - k - \chi\right)(t - m_{g, J}^2)} F(s, m_{g, J}^2, 4m_p^2 - s - m_{g, J}^2) \, .
\eeq
What we need is for the residue of this pole to be a polynomial whose leading term in $s$ is degree $J$, but the polynomial that arises from the ratios of Gamma functions doesn't quite do this, so we need the function $F$ to compensate.  The simplest way to allow this to happen is to make the crossing-symmetric form of $F(s, t, u)$ be
\beq
F(s, t, u) = \left(\frac{\alpha'_g s}{2}\right)^{2 - 2k} + \left(\frac{\alpha'_g t}{2}\right)^{2 - 2k} + \left(\frac{\alpha'_g u}{2}\right)^{2 - 2k} \, ,
\eeq
and require $k \le 1$, (so that this doesn't introduce additional poles into our amplitude).  Other, more exotic choices for $F$ clearly exist at this stage.  However, choosing $F$ to be polynomial in the Mandelstam variables is most consistent with the underlying idea that the basic form of closed string scattering is maintained even in a curved background; $F$ is then just a kinematic pre-factor, such as we know arises when we change the spins of the external string states even in bosonic string theory in 26-D flat space.

On the other hand, in the Regge limit our amplitude becomes 
\beq
\mathcal{A}_{g, \mathrm{Regge}} \approx \frac{C\pi \, e^{-i\pi\left(k - 1 + \frac{\alpha_g(t)}{2}\right)}\Gamma\left(1 - \frac{\alpha_g(t)}{2}\right)}{\Gamma\left(\frac{\alpha_g(t)}{2} - 1 - k - \chi\right)} \, \left(\frac{\alpha'_g s}{2}\right)^{\alpha_g(t)} \, ,
\eeq
which has exactly the scaling behavior we require.  This then gives the Reggeization prescription
\beq
\frac{1}{t - m_{g, 2}^2} \hspace{.25in} \rightarrow \hspace{.25in} \frac{\alpha' e^{-i\pi \left(k + \frac{\alpha_g(t)}{2}\right)} \, \Gamma\left(-k - \chi\right) \, \Gamma\left(1 - \frac{\alpha_g(t)}{2}\right)}{2 \, \Gamma\left(\frac{\alpha_g(t)}{2} - 1 - k - \chi\right)} \, \left(\frac{\alpha'_g s}{2}\right)^{\alpha_g(t) - 2} \, .
\eeq
Recall that in the standard prescription of \cite{DHM}, we have $k = 0$.  The integer $k$ appears in the above expression three times, but its presence in the phase factor is fairly trivial, generating at most a minus sign.  In the other two locations, we could interpret it as simply shifting the value of $\chi$ to $\chi + k$.  Usually we consider the value of $\chi$ fixed by the trajectory parameters and the mass of the proton, but this suggests from a purely phenomenological point of view one might interpret $\chi$ as an unknown parameter, to be determined via a data fitting scheme.  That being said, the choice $k = 0$ would still be the most consistent with the ideas that glueballs are dual to closed strings living in some curved spacetime background, and that the closed string amplitude in this background would retain most of the structure it has in flat space, with only the Regge trajectories changed.

\section{\label{mod2} A Second Possible Modification Scheme}

In order to further examine the role that $\chi$ plays in the Reggeization procedure, it is worth stepping back to the worldsheet integral that the Virasoro-Shapiro amplitude is derived from.  Let us first recall the argument in standard, 26-D flat space string theory.  We begin with four vertices, associated to tachyonic external string states, on a sphere.  The locations of three of the vertices can be fixed using conformal symmetry, leaving an integral over the fourth vertex location, which becomes an integral over the complex plane.  This is written
\beq
\label{eqn:wsintegral}
\mathcal{A}_{c} = C\int_{\mathbb{C}} d^2 z_4 \, |z_{12}|^2|z_{13}|^2|z_{23}|^2 \, \prod_{i < j} |z_{ij}|^{-\alpha' k_i\cdot k_j} \, ,
\eeq
where $\{z_1, z_2, z_3\}$ are the fixed locations of the first three vertices, $z_4$ is the location of the fourth, and $z_{ij} = z_i - z_j$.  It can be shown explicitly that the integral doesn't actually depend on the values of $\{z_1, z_2, z_3\}$.  Since rearranging the momenta $k_i$ is equivalent to rearranging the vertices $z_i$, this is how crossing symmetry manifests itself in this expression.   

The traditional method for solving this integral is to choose the values $\{0, 1, \infty\}$ for the first three vertex locations, giving
\beq
\mathcal{A}_{c} = C\int_{\mathbb{C}} d^2 z_4 \, |z_4|^{-4 - \frac{\alpha' u}{2}}|1 - z_4|^{-4 - \frac{\alpha' t}{2}} \, ,
\eeq
(where we have also rewritten the momentum dot products in terms of Mandelstam variables.)  This temporarily suppresses the crossing symmetry, but simplifies the integral so that it can be done in closed form, taking advantage of analytic continuation.  This gives the result
\beq
\mathcal{A}_{c} = \frac{C \, \pi \, \Gamma\left(-1 - \frac{\alpha' t}{4}\right) \, \Gamma\left(-1 - \frac{\alpha' u}{4}\right) \, \Gamma\left(3 + \frac{\alpha' t}{4} + \frac{\alpha' u}{4}\right)}{\Gamma\left(2 + \frac{\alpha' s}{4}\right) \, \Gamma\left(2 + \frac{\alpha' u}{4}\right) \, \Gamma\left(-2 - \frac{\alpha' t}{4} - \frac{\alpha' u}{4}\right)} \, ,
\eeq  
which can then be rewritten in a form where the crossing symmetry is manifest, using the mass-shell condition in equation \ref{eqn:BSMS}.  This results in the traditional form for the Virasoro-Shapiro amplitude, given in equation \ref{eqn:VSoriginal}.

Suppose instead of making modifications to the closed-form Virasoro-Shapiro amplitude, we go back to equation \ref{eqn:wsintegral}, and attempt to modify this.  This is subtle because the lack of dependence on the values $\{z_1, z_2, z_3\}$ relies on the conformal symmetry of the worldsheet, which is broken if we attempt to modify the mass shell condition and the Regge trajectory.\footnote{Presumably the true solution to this problem lies in properly quantizing strings on a curved background, to produce an exact dual to QCD.}  However, we note that if we choose $\{z_1, z_2, z_3\}$ to form an equilateral triangle, we retain explicit crossing symmetry.  Specifically, we choose
\beq
z_1 = e^{i\pi/3}, \hspace{.75in} z_2 = 0, \hspace{.75in} z_3 = 1 \, .
\eeq
Any translational shift or rotation of this triangle can be absorbed into a redefinition of the variable of integration, and any dilation of this triangle can be absorbed into a redefinition of the constant $C$; this is therefore a unique choice.  We then obtain
\beq
\mathcal{A}_{c} = C\int_{\mathbb{C}} d^2 z_4 \, |z_4|^{-4 - \frac{\alpha' t}{2}} \, |1 - z_4|^{-4 - \frac{\alpha' s}{2}} \, \left|e^{i\pi/3} - z_4\right|^{-4 - \frac{\alpha' u}{2}} \, .
\eeq

Now we replace the exponents $-4 - \frac{\alpha' x}{2}$ with an arbitrary linear function $B(x) = B_0 + B'x$, which we will later relate to the true Regge trajectory, and we allow for multiplying by an arbitrary function $\tilde{F}(s, t, u)$, yielding
\beq
\tilde{\mathcal{A}}_{g} = C\tilde{F}(s, t, u)\int_{\mathbb{C}} d^2 z_4 \, |z_4|^{-B(t)} \, |1 - z_4|^{-B(s)} \, \left|e^{i\pi/3} - z_4\right|^{-B(u)} \, .
\eeq
This amplitude is manifestly crossing symmetric, and we are assuming with an appropriate choice of $B_0$ and $B'$, it would have the correct pole structure and Regge limit to meet our requirements.  Following the traditional procedure, we should now compute this integral, confirm what the correct linear parameters are by examining the pole structure, and then take the Regge limit.  However, this integral is substantially more difficult, so we must work from the integral itself in examining both poles and the Regge limit.  We begin by using the physical mass-shell condition to rewrite our amplitude in terms of just $s$ and $t$, as
\beq
\label{eqn:stint}
\tilde{\mathcal{A}}_{g} = C\tilde{F}(s, t, 4m_p^2 - t - s)\int_{\mathbb{C}} d^2 z_4 \, |z_4|^{-B(t)} \, |1 - z_4|^{-B(s)} \, \left|e^{i\pi/3} - z_4\right|^{B(s) + B(t) - \chi_B} \, ,
\eeq
where $B(s) + B(t) + B(u) = \chi_B$.  

Our inability to perform this integral in closed form prevents us from examining the full pole structure at low energies, since this structure must arise from analytic continuation away from the region where the integral converges.  However, we can expand around the first pole $t \approx m_{g, 2}^2$, where $m_{g,2}$ is the mass of the spin-2 glueball.  We do this by noting that near the first pole, the integral must be dominated by the region where $z_4$ is small.  Using $z_4 = r e^{i\theta}$, and $\delta \ll 1$, this gives
\beq
\tilde{\mathcal{A}}_{g, t \approx m_{g, 2}^2} \approx 2\pi C\tilde{F}(s, m_{g, 2}^2, 4m_p^2 - m_{g, 2}^2 - s)\int_{0}^{\delta} r^{1 - B(t)} \, dr \approx \frac{2\pi C \delta^{2 - B(t)} \, \tilde{F}(s, m_{g, 2}^2, 4m_p^2 - m_{g, 2}^2 - s)}{2 - B(t)} \, ,
\eeq
which implies the first pole is at $B(t) = 2$.  That suggests we choose simply $B(t) = \alpha_g(t)$ (we will see that this is also supported by the Regge limit behavior), and this then gives
\beq
\tilde{\mathcal{A}}_{g, t \approx m_{g, 2}^2} \approx -\frac{2\pi C\tilde{F}(s, m_{g, 2}^2, 4m_p^2 - m_{g, 2}^2 - s)}{\alpha_g'(t - m_{g, 2}^2)} \, .
\eeq
Giving this pole the correct residue would then require that $\tilde{F}$ be quadratic in $s$.  Thus we choose
\beq
\tilde{F}(s, t, u) = \left(\frac{\alpha'_g s}{2}\right)^{2} + \left(\frac{\alpha'_g t}{2}\right)^{2} + \left(\frac{\alpha'_g u}{2}\right)^{2} \, .
\eeq 

Next we apply the Regge limit directly to equation \ref{eqn:stint}.  This integral does not converge for large real values of $s$, so in order to perform it in the Regge limit, we will allow $s$ to have a large imaginary part, and analytically continue back to physical values of $s$ after integration.  With $s$ large and complex, the integrand is largest for $z_4 \sim \frac{1}{s}$, close to the origin.  We can therefore write
\beq
|1 - z_4|^{-B(s)} \approx e^{\frac{B's}{2}(z_4 + \bar{z}_4)}, \hspace{1in} \left|e^{i\pi/3} - z_4\right|^{B(s) + B(t) - \chi_B} \approx e^{-\frac{B's}{2}\left(z_4 e^{-i\pi/3} + \bar{z}_4 e^{i\pi/3}\right)} \, ,
\eeq
so that our integral becomes
\beq
\tilde{\mathcal{A}}_{g, \mathrm{Regge}} \approx C\left(\frac{\alpha'_g s}{2}\right)^{2}\int_{\mathbb{C}} d^2 z_4 \, |z_4|^{-B(t)} \, e^{\frac{B's}{2}\left(z_4 - z_4e^{-i\pi/3} + \bar{z}_4 - \bar{z}_4 e^{i\pi/3}\right)} \, ,
\eeq
which yields
\beq
\tilde{\mathcal{A}}_{g, \mathrm{Regge}} \approx \frac{C\pi \, e^{-i\pi\left(\frac{B(t)}{2} - 1\right)} \, \Gamma\left(1 - \frac{B(t)}{2}\right)}{\Gamma\left(\frac{B(t)}{2}\right)} \, \left(\frac{B's}{2}\right)^{B(t) - 2}\left(\frac{\alpha'_g s}{2}\right)^{2} \, .
\eeq
Again we see that in order for this to have the correct poles and scaling behavior, we must choose $B(t) = \alpha_g(t)$, which then gives
\beq
\tilde{\mathcal{A}}_{g, \mathrm{Regge}} \approx \frac{C\pi \, e^{-i\pi\left(\frac{\alpha_g(t)}{2} - 1\right)} \, \Gamma\left(1 - \frac{\alpha_g(t)}{2}\right)}{\Gamma\left(\frac{\alpha_g(t)}{2}\right)} \, \left(\frac{\alpha'_gs}{2}\right)^{\alpha_g(t)} \, .
\eeq
Comparison between this result and the pole expansion then leads to the Reggeization prescription
\beq
\frac{1}{t - m_{g, 2}^2} \hspace{.25in} \rightarrow \hspace{.25in} \frac{\alpha_g' \, e^{-\frac{i\pi\alpha_g(t)}{2}} \, \Gamma\left(1 - \frac{\alpha_g(t)}{2}\right)}{2\Gamma\left(\frac{\alpha_g(t)}{2}\right)} \, \left(\frac{\alpha'_gs}{2}\right)^{\alpha_g(t) - 2} \, .
\eeq

This is very similar to the solution found in \cite{DHM}, but no parameter $\chi$ appears in the final result.  Equivalently, you could say it takes the same form as the original solution but with $\chi = -1$.  This reinforces the conclusion of the previous section, that choosing a Reggeization procedure inspired by the structure of closed string scattering in 26-D flat space is not completely unique.  We will therefore assume a generic form
\beq
\frac{1}{t - m_{g, 2}^2} \hspace{.25in} \rightarrow \hspace{.25in} \frac{\alpha_g' \, e^{-\frac{i\pi\alpha_g(t)}{2}} \, \Gamma(-\chi)\Gamma\left(1 - \frac{\alpha_g(t)}{2}\right)}{2\Gamma\left(\frac{\alpha_g(t)}{2} - 1 - \chi\right)} \, \left(\frac{\alpha'_gs}{2}\right)^{\alpha_g(t) - 2} \, ,
\eeq
with $\chi$ undertermined, and we will use this in fitting elastic proton-proton scattering.  We can then examine what value of $\chi$ agrees best with the data, and use this as a guide in evaluating which modification scheme ought to be used.

\section{\label{fitting} Fitting to Proton-Proton Scattering with an Additional Free Parameter}

\subsection{\label{fitfunction} The Differential Cross Section}

\begin{figure}
\begin{center}
\resizebox{2in}{!}{\includegraphics{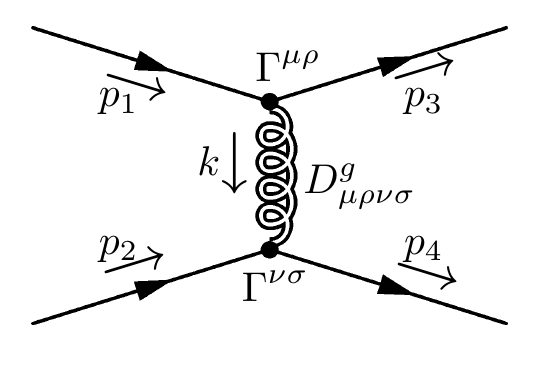}}
\caption{\label{feyn} The Feynman diagram for proton-proton scattering via tree-level glueball exchange in the $t$-channel.}
\end{center}
\end{figure}

In order to use data to evaluate how well different values of $\chi$ work, we must use this Reggeization procedure to model proton-proton scattering via Pomeron exchange.  This begins by calculating the amplitude and differential cross section for the scattering process via the exchange of a spin-2 massive glueball in the Regge limit, for which the Feynman diagram is shown in FIG. \ref{feyn}.  The propagator for a massive spin-2 particle is 
\beq
D_{\mu\rho\nu\sigma} = \frac{\frac{1}{2}(\eta_{\mu\nu}\eta_{\rho\sigma} + \eta_{\mu\sigma}\eta_{\rho\nu}) + \cdots}{t - m_{g,2}^2} \, ,
\eeq
where the terms not written will either vanish when contracted into the vertices or be suppressed in the Regge limit \cite{prop}.  The vertex structures we will use are
\beq
\Gamma^{\mu\rho}(P_1) = \frac{i\lambda A(t)}{2}\left(\gamma^{\mu}P_1^{\rho} + \gamma^{\rho}P_1^{\mu}\right) \ + \dots \, ,
\eeq
where $P_1 = \frac{p_1 + p_3}{2}$.  (We will similarly define $P_2 = \frac{p_2 + p_4}{2}$.)  This vertex structure is based on assuming the glueballs couple to the protons predominately via the QCD stress tensor, and we have again ignored terms that will not contribute significantly in the Regge limit \cite{Domokos:2010ma}.  Finally, $A(t)$ is a form factor that should be well approximated by a dipole form
\beq
A(t) = \frac{1}{\left(1 - \frac{t}{M_d^2}\right)^2} \, ,
\eeq
for the values of $t$ we are considering \cite{Hong:2007dq}.  

Putting these pieces together the amplitude for the process is
\beq
\mathcal{A} = \Big[\bar{u}_3\Gamma^{\mu\rho}(P_1)u_1\Big]D_{\mu\rho\nu\sigma}(k)\Big[\bar{u}_4\Gamma^{\nu\sigma}(P_2)u_2\Big] \, ,
\eeq
which in the Regge limit leads to
\beq
\frac{1}{4}\sum_{\mathrm{spins}} |\mathcal{A}|^2 = \frac{\lambda^4 \, A^4(t) \, s^4}{(t - m_{g, 2}^2)^2} \, .
\eeq
If we replace the propagators with our Reggeized propagators, and use this expression to find the differential cross section, we obtain
\beq
\label{model}
\frac{d\sigma}{dt} = \frac{\lambda^4 A^4(t) \Gamma^2(-\chi)\Gamma^2\left(1 - \frac{\alpha_g(t)}{2}\right)}{16\pi \Gamma^2\left(\frac{\alpha_g(t)}{2} - 1 - \chi\right)} \,\left(\frac{\alpha_g' s}{2}\right)^{2\alpha_g(t) - 2} \, .
\eeq

\subsection{The Data Fitting Results}

We now want to fit this model to existing proton-proton and proton-antiproton scattering data.  We will restrict our attention to scattering processes where single soft Pomeron exchange is the (presumed) dominate contributor.  Based on FIG. \ref{totalcross}, we will consider only data with $\sqrt{s} > 500$ GeV, where the contribution from Reggeon exchange is less than 1\%.  We also restrict ourselves to the range $0.01 < |t| < 0.6$ GeV; below $|t| = 0.01$ GeV there are significant Coulomb interactions, and above $|t| > 0.6$ GeV we are in the hard Pomeron regime.  This leaves us with three available center-of-mass energies: $\sqrt{s} = 546$ GeV and $\sqrt{s} = 1800$ GeV, from the E710 and CDF experiments at the Tevatron, and $\sqrt{s} = 7$ TeV, from the TOTEM experiment at the LHC.  This data was taken from the High Energy Physics Data Repository (https://hepdata.net).

\begin{figure}
\begin{center}
\resizebox{4in}{!}{\includegraphics{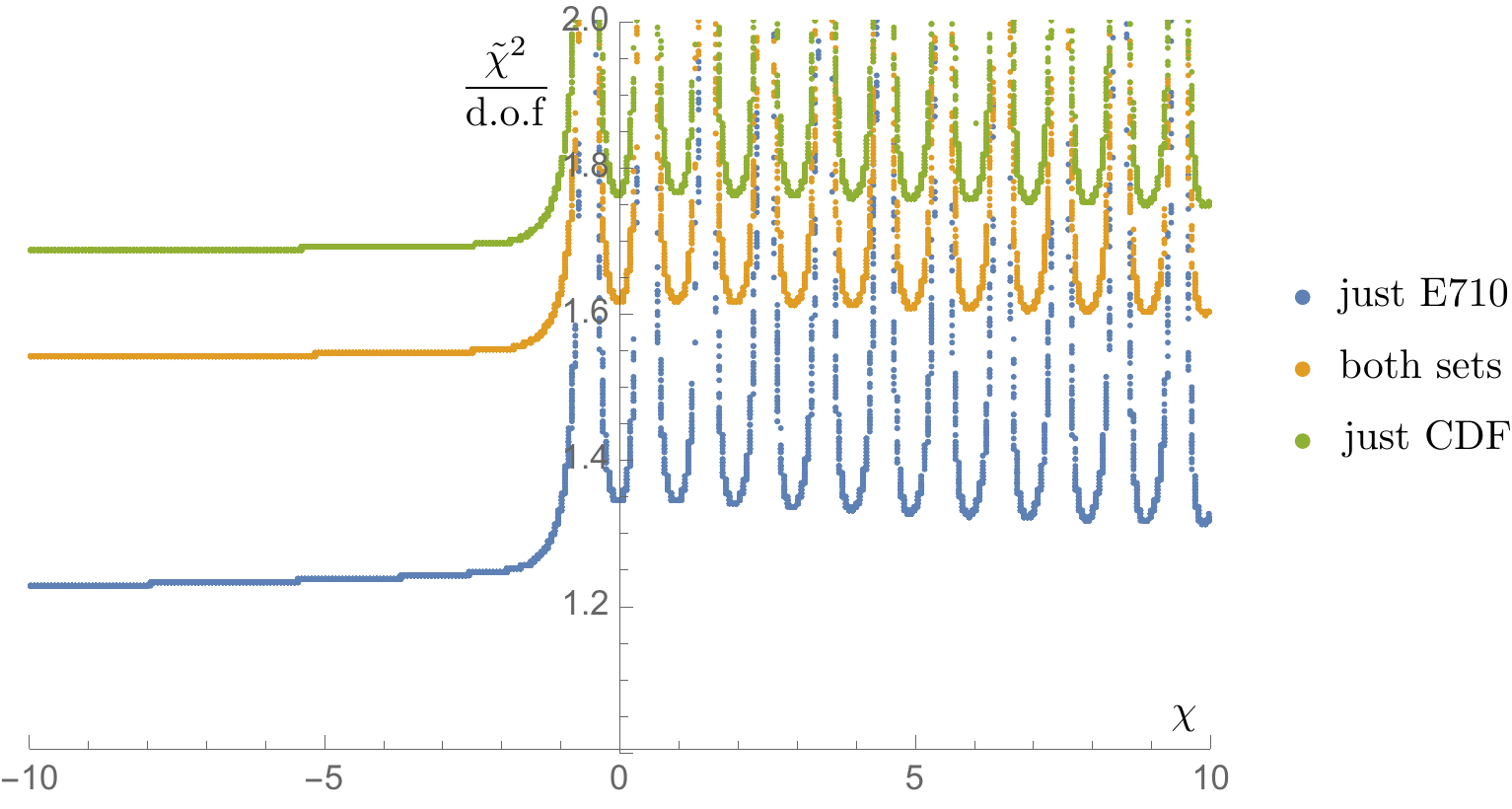}}
\caption{\label{xivschi} A map of the minimum value of $\frac{\tilde{\chi}^2}{\mathrm{d.o.f}}$ with respect to the other fitting parameters, for each value of $\chi$}
\end{center}
\end{figure}

We will treat equation \ref{model} as our model, with five fitting parameters $\{\lambda, \alpha_{g0}, \alpha_{g}', M_d, \chi\}$, and use a standard weighted fit in which we minimize the quantity
\beq
\frac{\tilde{\chi}^2}{\mathrm{d.o.f}} = \frac{1}{\mathrm{d.o.f}}\sum_{\mathrm{data \ points}} \frac{1}{\tilde{\sigma}^2_{\mathrm{exp}}}\left[\left(\frac{d\sigma}{dt}\right)_{\mathrm{exp}} - \left(\frac{d\sigma}{dt}\right)_{\mathrm{model}}\right]^2
\eeq
where $\mathrm{d.o.f}$ is the number of degrees of freedom in the fit, $\tilde{\sigma}_{\mathrm{exp}}$ is the experimental uncertainty in each data point, $\left(\frac{d\sigma}{dt}\right)_{\mathrm{exp}}$ is the experimental value of the differential cross section at each data point, and $\left(\frac{d\sigma}{dt}\right)_{\mathrm{model}}$ is the value of the differential cross section that the model provides at each data point.\footnote{Unfortunately, the traditional notation for discussing data fitting overlaps with the notation used elsewhere in this discussion, so the tildes are added for clarity.}  

Although the value of $\frac{\tilde{\chi}^2}{\mathrm{d.o.f}}$ will be our primary measure for evaluating the fit, and our primary interest is in exploring various choices of the parameter $\chi$, it is also helpful to keep in mind information we have about the values of the other fitting parameters.  As was discussed in section \ref{PomeronReview}, analysis of total cross sections for different center-of-mass energies suggests $\alpha_{g0} \approx 1.085$.  The slope of the Pomeron trajectory based on similar analyses is usually given somewhere around $\alpha_{g}' \approx 0.3 \ \mathrm{GeV}^{-2}$ \cite{CGM}.  The values of $M_d$ and $\lambda$ are less well established, but they can be computed in AdS/QCD dual models; a Skyrme model for the proton generates $M_d = 1.17 \ \mathrm{GeV}$ and $\lambda = 9.02 \ \mathrm{GeV}^{-1}$ \cite{DHM, Domokos:2010ma}.

A simple automated approach to this fitting problem encounters issues associated with the dependence on $\chi$.  It appears as an argument of the Gamma function, and the regularly spaced poles of this function produce many possible fit values for $\chi$, because the quantity $\frac{\tilde{\chi}^2}{\mathrm{d.o.f}}$ has a series of local minima in the variable $\chi$.  This complicates the fit, so our approach is to fix values of $\chi$ and fit with respect to the other parameters, then extract the values of $\frac{\tilde{\chi}^2}{\mathrm{d.o.f}}$ for each.  The result is a map such as that shown in FIG. \ref{xivschi}. 

\begin{figure}
\begin{center}
\resizebox{4in}{!}{\includegraphics{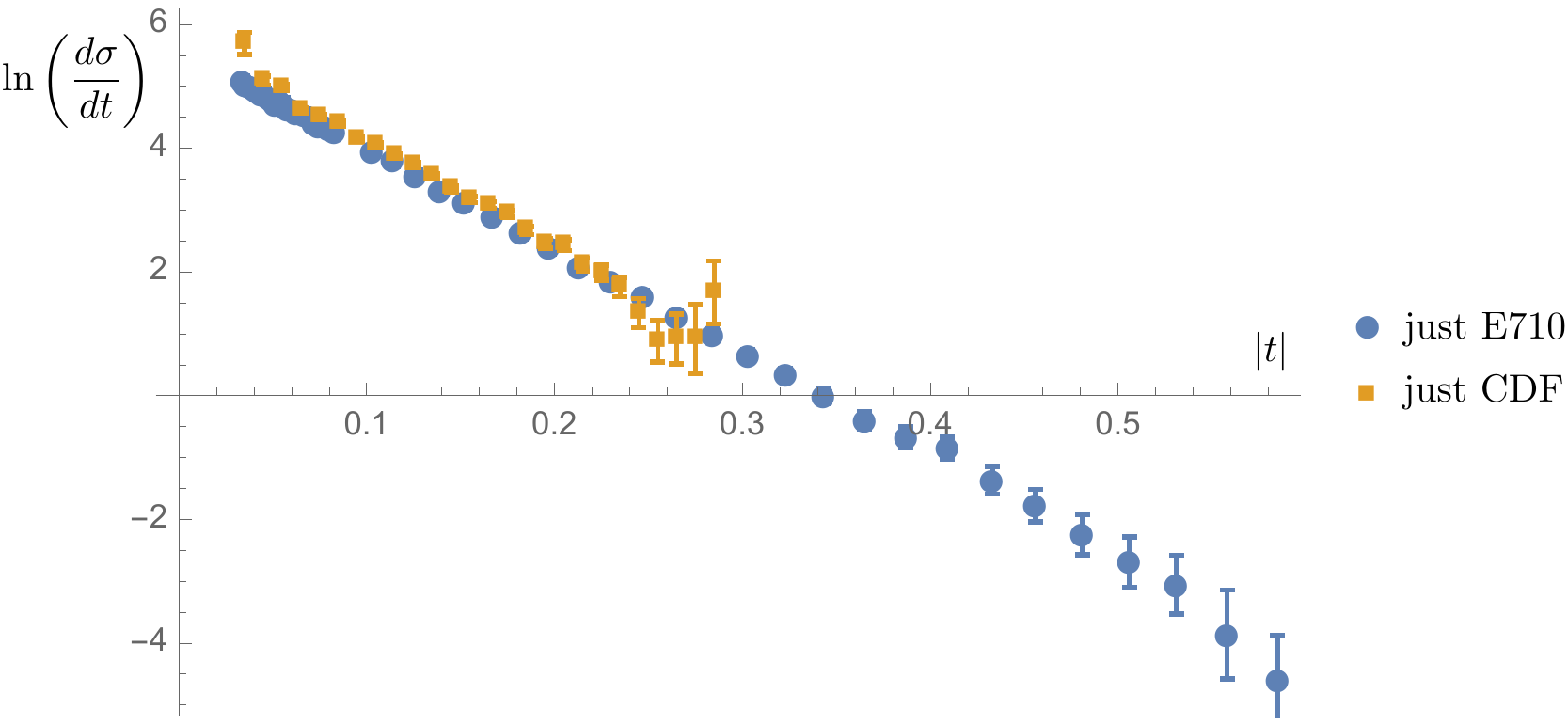}}
\caption{\label{datadiscrepancy} The discrepancy between the CDF and E710 data at $\sqrt{s} = 1800$ GeV}
\end{center}
\end{figure}

There is a known discrepancy at the energy 1800 GeV between the data produced by the E710 experiment and that produced by the CDF experiment, as shown in FIG. \ref{datadiscrepancy}, and discussed in \cite{DHM}.  In that work removing either data set from the analysis resulted in a better fit, with a somewhat better result when just the CDF data remained.  FIG. \ref{xivschi} includes maps of the best fit value of $\frac{\tilde{\chi}^2}{\mathrm{d.o.f}}$ as functions of $\chi$, for data including each experiment as well as including both.  Interestingly, here we see the opposite from what was previously found: the fit is markedly better if we only include the data from the E710 experiment.  In fact, having both data sets involved is still better than just involving the CDF data.  Since this is true systematically, regardless of the value of $\chi$, we conclude that this is largely because, at least assuming our model, the E710 experimental data at 1800 GeV is more consistent with the data at the other center-of-mass energies, so we will continue our analysis excluding the CDF data set at 1800 GeV.

Further considering FIG. \ref{xivschi}, there is a series of (close to) regularly spaced local minima mostly in the region of positive $\chi$.  These minima get slightly lower as $\chi$ increases, but all give very similar values of $\frac{\tilde{\chi}^2}{\mathrm{d.o.f}} \approx 1.34$.  They are also all associated to similar values for the other fitting parameters.  Since none of the theoretical models suggest values of $\chi$ in the range $\chi > 1$, we will focus on the first two local minima.  These fits are shown in table \ref{fittable}.  Note that the values of the other fitting parameters differ significantly from their predicted values.  Furthermore, these local minima are associated to values of $\frac{\tilde{\chi}^2}{\mathrm{d.o.f}}$ that are somewhat higher than the typical value of $\frac{\tilde{\chi}^2}{\mathrm{d.o.f}}$ for negative $\chi$.

\begin{table}
\begin{center}
\begin{tabular}{|c|c|c|c|}
\hline
\ \ fit parameters \ \ & \ \ first local minimum \ \ & \ \ second local minimum \ \ & $\chi = -1$ \\
\hline 
\hline
$\lambda$ & $1.608 \pm 0.001$ & $1.881 \pm 0.001$ & $6.729 \pm 0.004$ \\
$\alpha_{g0}$ & $1.0964 \pm 0.0001$ & $1.0961 \pm 0.0001$ & \ \ $1.09472 \pm 0.00006$ \ \ \\
$\alpha_{g}'$ & $0.570 \pm 0.001$ & $0.566 \pm 0.001$ & $0.5475 \pm 0.0009$ \\
$M_d$ & $3.87 \pm 0.10$ & $2.58 \pm 0.03$ & $1671 \pm 91000$ \\
$\chi$ & $-0.02585 \pm 0.00003$ & $0.94977 \pm 0.00006$ & -1 \\
\hline
\hline
$\frac{\tilde{\chi}^2}{\mathrm{d.o.f}}$ & 1.344 & 1.343 & 1.336 \\
\hline
\end{tabular}
\caption{\label{fittable} A table of fitting results associated with fixed values of $\chi$.}
\end{center}
\end{table} 
 
For most of the negative regime the map is close to flat, with a lower value of $\frac{\tilde{\chi}^2}{\mathrm{d.o.f}}$, but the fit shows a slight preference for increasingly negative values.  In the limit that $\chi$ is large and negative, we are effectively working with a different, simpler model:
 \beq
 \frac{d\sigma}{dt}\Bigg|_{\chi \ll 0} \approx \frac{\lambda^4 A^4(t)}{16\pi}\Gamma^2\left(1 - \frac{\alpha_g(t)}{2}\right) \, (-\chi)^{1 - \frac{\alpha_g(t)}{2}} \, \left(\frac{\alpha' s}{2}\right)^{2\alpha_g(t) - 2} \, .
 \eeq 
One might conclude that this model is a better fit for the data.  However, the values of $\frac{\tilde{\chi}^2}{\mathrm{d.o.f}}$ that result from this fit are not significantly lower than those for more moderate, negative values of $\chi$.  If we decide based on FIG. \ref{xivschi} that the best fit is for some negative value of $\chi$, we still cannot really argue that any particular negative value should be chosen.

\begin{figure}
\begin{center}
\resizebox{4in}{!}{\includegraphics{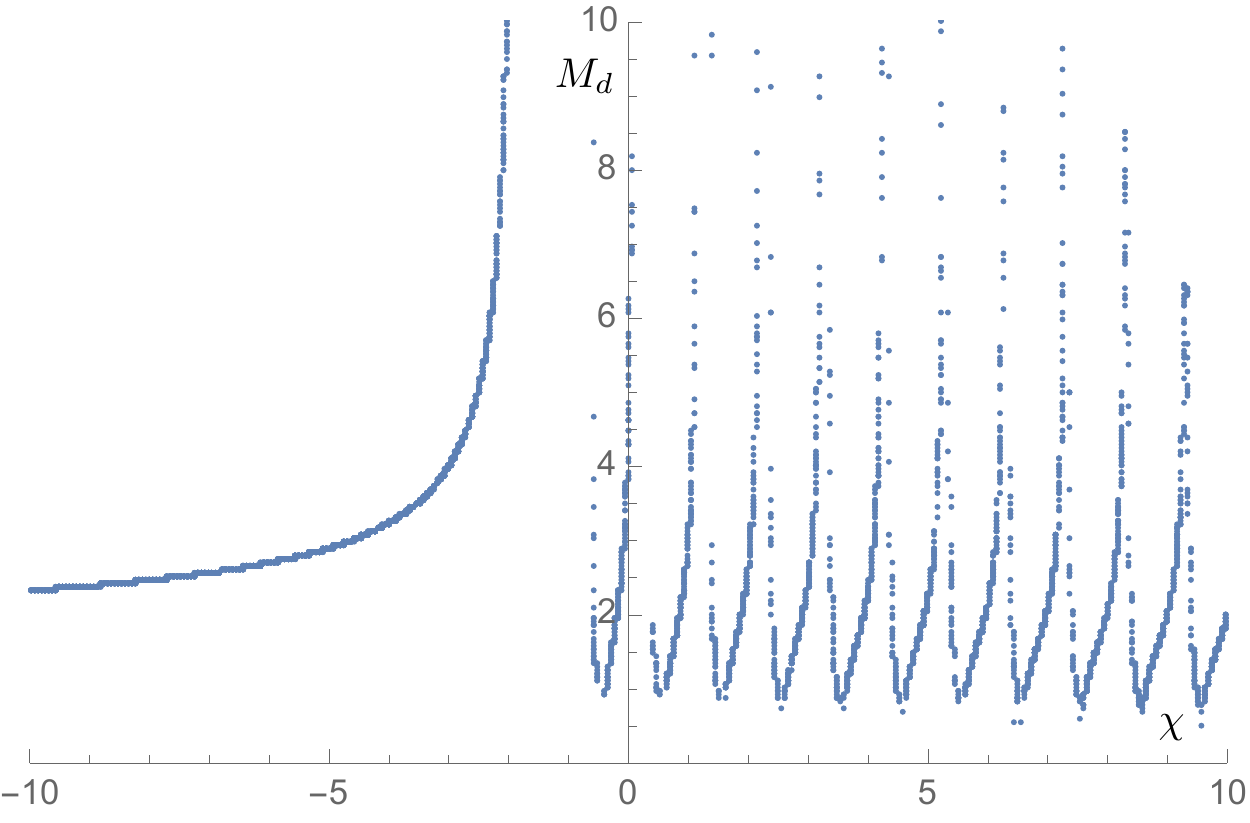}}
\caption{\label{chivsmd} A map of the fit value of $M_d$, for each value of $\chi$}
\end{center}
\end{figure}

That being said, for a region of moderate negative values of $\chi \in [-1.942, -0.615]$, the fit attempts result in ``runaway'' values of the fitting parameter $M_d$: the fit chooses an extremely large value of $M_d$, with an even larger uncertainty, which suggests that this range of $\chi$ values should be ruled out.  See for example a plot of the fit dipole masses associated with each value of $\chi$, as shown in FIG. \ref{chivsmd}.  This behavior is exhibited in particular by the value $\chi = -1$, the value suggested by the modification scheme developed in section \ref{mod2}.  We have included these fitting results in table \ref{fittable}.

Although the general data fitting is not pointing to any particular value of $\chi$, we can still examine specifically the family of possible choices of $\chi$ suggested in section \ref{mod1}, which is
\beq
\label{chik}
\chi = \frac{3\alpha_{g0}}{2} - 3 + 2\alpha_g' M_p^2 + k, \hspace{1in} k = 1, 0, -1, -2, \dots \, .
\eeq
The first several choices of $k$ are shown in table \ref{fittable2}.  Of primary interest is the choice $k = 0$, which corresponds to the original value of $\chi$ used in \cite{DHM}.  This generates a fit with a value of $\frac{\tilde{\chi}^2}{\mathrm{d.o.f}}$ that is around twice that associated with the lowest values we have.  However, the values of the other fitting parameters are much closer to their predicted values in this case than for any other choices we are considering.

The choices $k = \pm 1$ generate fits with the runaway behavior in the fitting parameter $M_d$.  However, choices $k = -2, -3, -4, \cdots$ work well.  These generate fits consistent with the generic behavior for negative $\chi$: they all have $\frac{\tilde{\chi}^2}{\mathrm{d.o.f}} \approx 1.2$, with a slight improvement in the fit as $k$ decreases.  On the other hand, the other fitting parameters are then significantly different from their predicted values.

\begin{table}
\begin{center}
\begin{tabular}{c|c|c|c|c|}
fit parameters & $k = 1$ & $k = 0$ & $k = -1$ & $k = -2$ \\
\hline 
$\lambda$ & $0.05981 \pm 0.00004$ & $10.930 \pm 0.007$ & $5.703 \pm 0.004$ & $4.637 \pm .003$ \\
$\alpha_{g0}$ & $1.12531375 \pm 6 \times 10^{-8}$ & $1.0834 \pm 0.0001$ & $1.09626 \pm 0.00009$ & $1.09734 \pm 0.00009$ \\
$\alpha_{g}'$ & $0.74564453 \pm 5 \times 10^{-8}$ & $0.4173 \pm 0.0001$ & $0.5665 \pm 0.0005$ & $0.5798 \pm 0.0007$ \\
$M_d$ & $79390 \pm 3 \times 10^{7}$ & $1.91 \pm 0.01$ & $103876 \pm 6 \times 10^8$ & $5.7 \pm 0.3$ \\
\hline
$\frac{\tilde{\chi}^2}{\mathrm{d.o.f}}$ & 24.5 & 2.63 & 1.269 & 1.246 \\
\end{tabular}
\caption{\label{fittable2} A table of fitting results assuming the family of choices for $\chi$ discussed in section \ref{mod1}.}
\end{center}
\end{table}

\begin{figure}
\begin{center}
\resizebox{4in}{!}{\includegraphics{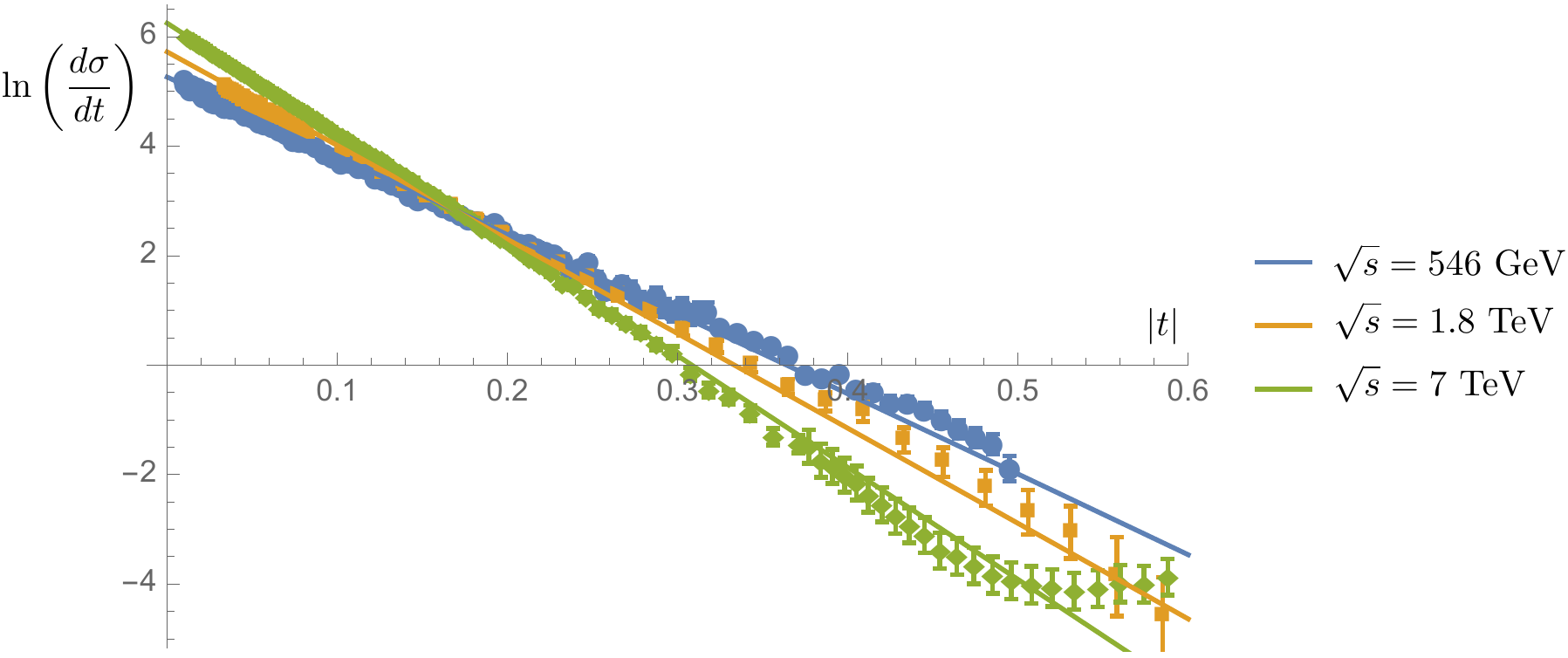}}
\caption{\label{fitresult} The data (displayed on a log plot) together with the model for the fitting results using equation \ref{chik} with $k = -2$.}
\end{center}
\end{figure}

The data and the fit are shown together for the choice $k = -2$ in FIG. \ref{fitresult}.  What we notice here is that the fit is best for small values of $|t|$, with deviations mostly in the range $0.5 \ \mathrm{GeV}^2 < |t| < 0.6 \ \mathrm{GeV}^2$.  These deviations are most pronounced for the $\sqrt{s} = 7$ TeV data, though they are also somewhat apparent for the lower energy data.  This might indicate that the transition between hard Pomeron and soft Pomeron behaviors occurs at $t = -0.5$ GeV$^2$ rather than $t = 0.6 \ \mathrm{GeV}^2$.

In summary, it seems that there are two possible conclusions based on our analysis.  The best way to minimize $\frac{\tilde{\chi}^2}{\mathrm{d.o.f}}$ is to choose a value of $\chi < -1.942$.  However, no particular value is strongly better than any other, and the other fitting parameters do not agree with our predictions.  On the other hand, we could choose the traditional value of $\chi$ first given in \cite{DHM}.  This results in a significantly weaker fit, but fitting parameters (other than $\chi$) that are more reasonable.  

\section{\label{Conclusion} Conclusions and Future Directions}

In this work we have explored the role of the mass-shell parameter $\chi$ in the Reggeization procedure modeling proton-proton scattering via Pomeron exchange.  The essential idea behind our model is to start with the cross section for proton-proton scattering via the lowest lying particle on the Pomeron trajectory, the spin 2 glueball.  Then, we replace the propagator with a ``Reggeized'' propagator, that should take into account the exchanges of all the particles in the trajectory.  This replacement rule is based on analyzing the Virasoro-Shapriro amplitude for the scattering of four closed strings and modifying the amplitude to depend on the physical Pomeron trajectory.

The original method of Reggeization was developed in \cite{DHM}, and involved the introduction of the parameter $\chi$, based on the mass-shell condition for the protons.  However, in sections \ref{mod1} and \ref{mod2} we showed that this method could be generalized and modified while still satisfying the basic phenomenological requirements of proton-proton scattering via Pomeron exchange (reviewed in section \ref{PomeronReview}).  These modifications effectively change the value of $\chi$.  In order to better inform the choice of $\chi$ and compare the effectiveness of the original scheme with the generalized one we fit the model to proton-proton scattering data in section \ref{fitting}, allowing $\chi$ to be a fitting parameter.  

The fitting procedure was complicated by the role the parameter $\chi$ plays in the model; its appearance inside a gamma function leads to a landscape of fitting results with multiple locally ``best fit'' choices.  We therefore performed the fit by choosing values of $\chi$ and fitting to the other parameters, thus creating a map of the best possible $\frac{\tilde{\chi}^2}{\mathrm{d.o.f}}$ for values of $\chi$ in the range $[-10, 10]$.  We also analyzed the specific choices of $\chi$ suggested in sections \ref{mod1} and \ref{mod2}.  The results of this analysis were inconclusive: the smallest values of $\frac{\tilde{\chi}^2}{\mathrm{d.o.f}}$ are acheived for $\chi < -1.942$.  However, the landscape of $\frac{\tilde{\chi}^2}{\mathrm{d.o.f}}$ is too flat to effectively narrow down the choice of $\chi$ beyond that, and the other fitting parameters in this region seem inconsistent with what we know about them.  On the other hand, the original choice of $\chi$ given in \cite{DHM} provides a substantially higher value of $\frac{\tilde{\chi}^2}{\mathrm{d.o.f}}$, but generates values for the other parameters that are a better fit to previous work.  We \emph{were} able to rule out the modification from section \ref{mod2} with reasonable certainty.

One insight that we gained involved the experimental data at center-of-mass energy $1800 \ \mathrm{GeV}$.  As has been previously established, there is some discrepancy between the data sets obtained by the experiments E710 and CDF at the Tevatron.  Our model systematically fits the E710 data better than the CDF data for any choice of $\chi$; in fact, the CDF data alone generates a worse fit than even both data sets together.  This result is different than that found in \cite{DHM}; it seems likely that the inclusion in our fitting of the 7 TeV data from the TOTEM experiment at the LHC is the source of this change: our model suggests that the E710 data is more consistent with the new 7 TeV data.  And while this might be model specific, the most important factor in how the data sets at different values of $\sqrt{s}$ relate to each other is the scaling $\frac{d\sigma}{dt} \propto s^{2\alpha_g(t) - 2}$, which is a common feature of any Pomeron exchange model.  The fitting results also showed the greatest discrepancies with the model in the range $0.5 \ \mathrm{GeV}^2 < |t| < 0.6 \ \mathrm{GeV}^2$, which might suggest that the transition from soft Pomeron to hard Pomeron behavior occurs at a different location than previously thought: $t = -0.5 \ \mathrm{GeV}^2$.  

In the future, it might be interesting to repeat this analysis over a larger range of values of $\sqrt{s}$.  We could incorporate lower energy scattering data effectively if we modified our model to include Reggeon exchange as well as Pomeron exchange.  To include higher energy data we would need to wait for the LHC to provide results at $14 \ \mathrm{TeV}$.  However, this might give us additional clarity on the issue of the discrepancy between the two 1800 GeV data sets, and that of the deviations between the model and the fit for $0.5 \ \mathrm{GeV}^2 < |t| < 0.6 \ \mathrm{GeV}^2$.

\begin{acknowledgements}
Z. Hu and N. Mann would like to acknowledge the support of the Union College Summer Research Fellowship program.
\end{acknowledgements}

\end{document}